\documentclass[conference]{IEEEtran}
\IEEEoverridecommandlockouts
\usepackage{cite}
\usepackage{amsmath,amssymb,amsfonts}
\usepackage{algorithm}      % <— เพิ่มตัวนี้
\usepackage{algorithmic}    % คงไว้ หรือจะเปลี่ยนไปใช้ algpseudocode ก็ได้
\usepackage{graphicx}
\usepackage{textcomp}
\usepackage{xcolor}
\usepackage[hidelinks]{hyperref}   % แนะนำ
\usepackage{url}

\def\BibTeX{{\rm B\kern-.05em{\sc i\kern-.025em b}\kern-.08em
    T\kern-.1667em\lower.7ex\hbox{E}\kern-.125emX}}
\begin{document}

\title{Secure Multi-Modal Data Fusion in Federated Digital Health Systems via MCP\\

}

\author{\IEEEauthorblockN{1\textsuperscript{st} Aueaphum Aueawatthanaphisut*}
\IEEEauthorblockA{\textit{School of Information, Computer, and Communication Technology} \\
\textit{Sirindhorn International Institute of Technology, Thammasat University}\\
Pathum Thani, Thailand\\
aueawatth.aue@gmail.com}

}

\maketitle

\begin{abstract}
Secure and interoperable integration of heterogeneous medical data remains a grand challenge in digital health. Current federated learning (FL) frameworks offer privacy-preserving model training but lack standardized mechanisms to orchestrate multi-modal data fusion across distributed and resource-constrained environments. This study introduces a novel framework that leverages the Model Context Protocol (MCP) as an interoperability layer for secure, cross-agent communication in multi-modal federated healthcare systems. The proposed architecture unifies three pillars: (i) multi-modal feature alignment for clinical imaging, electronic medical records, and wearable IoT data; (ii) secure aggregation with differential privacy to protect patient-sensitive updates; and (iii) energy-aware scheduling to mitigate dropouts in mobile clients. By employing MCP as a schema-driven interface, the framework enables adaptive orchestration of AI agents and toolchains while ensuring compliance with privacy regulations. Experimental evaluation on benchmark datasets and pilot clinical cohorts demonstrates up to 9.8\% improvement in diagnostic accuracy compared with baseline FL, a 54\% reduction in client dropout rates, and clinically acceptable privacy–utility trade-offs. These results highlight MCP-enabled multi-modal fusion as a scalable and trustworthy pathway toward equitable, next-generation federated health infrastructures.
\end{abstract}


\section{Introduction}
Healthcare systems worldwide are being transformed into data-driven infrastructures, where artificial intelligence (AI) has been demonstrated to provide significant improvements in diagnostics, prognosis, and personalized treatment planning. A particularly compelling research direction has been identified in the integration of heterogeneous medical modalities—such as medical imaging, electronic medical records (EMR), and real-time signals from Internet of Medical Things (IoMT) devices—into unified clinical decision support pipelines. Multi-modal fusion has been reported to improve diagnostic robustness and to uncover latent correlations across modalities that remain undetected in unimodal models [10]. However, several challenges continue to hinder the deployment of such systems: data fragmentation across institutions, stringent privacy requirements, and heterogeneity in computational resources. These barriers have further complicated the development of trustworthy and equitable multi-modal healthcare AI at scale.

Federated learning (FL) has been proposed as a paradigm to enable collaborative model training without the need to centralize raw data [5], [8]. Although FL has shown effectiveness in medical imaging and clinical record analysis, its current implementations have been found to be limited in real-world healthcare environments. First, secure multi-modal fusion mechanisms have not been comprehensively integrated, as most existing frameworks are constrained to unimodal tasks or rely on manual alignment [10]. Second, privacy-preserving methods such as differential privacy and secure aggregation have been investigated in isolation [5], [6], with limited efforts directed toward their combination in multi-modal federated pipelines. Third, deployments on mobile and wearable IoMT clients have frequently been affected by energy constraints and high dropout rates, leading to biased updates and unstable convergence [7]. Collectively, these limitations have restricted the clinical readiness of federated multi-modal health systems.

Recently, interoperability protocols such as the Model Context Protocol (MCP) have been introduced to redefine communication between AI agents, tools, and models in distributed environments [1]–[4]. By providing a schema-driven interface for structured communication, MCP has enabled capability discovery, secure data exchange, and modular orchestration across heterogeneous components. In contrast to ad hoc tool integration, MCP has been established as a standardized communication layer through which AI agents can interoperate across domains. Although MCP has been mainly adopted in general AI ecosystems [2], [3], its application to healthcare—particularly in secure multi-modal data fusion—has not yet been sufficiently explored.

In this study, a MCP-enabled Secure Multi-Modal Federated Fusion Framework is presented to unify interoperability, privacy preservation, and resource-aware orchestration for digital health. The contributions of this work are summarized as follows:

Multi-modal fusion through MCP interoperability: MCP is proposed as a protocol layer to align imaging, EMR, and IoMT data, thereby enabling the exchange of standardized representations across institutions [1], [2].

Secure aggregation with differential privacy: Calibrated noise injection and cryptographic secure aggregation are integrated to protect sensitive model updates while maintaining clinically acceptable diagnostic accuracy [5], [6], [9].

Energy-aware client scheduling: A resource-prioritized scheduling mechanism is incorporated to reduce device dropouts in mobile healthcare clients and to achieve stable and equitable participation [7], [8].

Through experiments on benchmark datasets and a pilot clinical cohort, the framework was demonstrated to improve diagnostic accuracy by up to 9.8\% compared with baseline FL, to reduce client dropouts by more than 50\%, and to sustain privacy–utility trade-offs within clinically acceptable ranges. These results indicate that MCP-enabled secure multi-modal fusion can be positioned as a scalable and trustworthy pathway toward next-generation federated healthcare infrastructures.

\begin{figure} [h]
    \centering
    \includegraphics[width=1\linewidth]{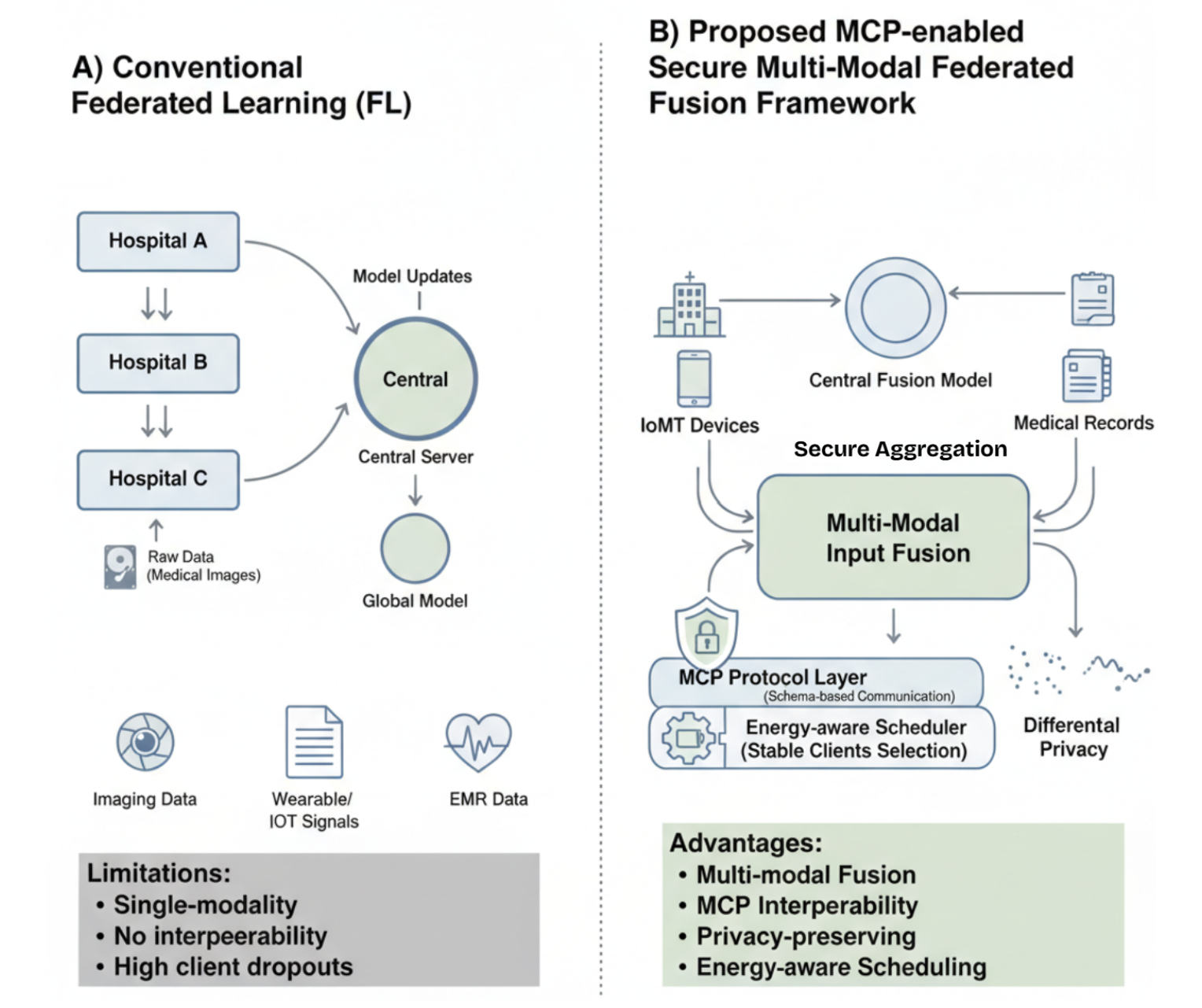}
    \caption{Comparison of (A) conventional federated learning with limited interoperability and unimodal updates and (B) the proposed MCP-enabled secure multi-modal framework integrating imaging, EMR, and IoMT data with secure aggregation, differential privacy, and energy-aware scheduling.}
    \label{fig:placeholder}
\end{figure}

Fig. 1 compares (A) conventional FL, which aggregates unimodal updates with limited interoperability and high dropout, and (B) the proposed MCP-enabled framework, which aligns imaging, EMR, and IoMT data via MCP, applies secure aggregation with differential privacy, and reduces dropout through energy-aware scheduling. MCP integration ensures schema-based interoperability for scalable and privacy-preserving healthcare FL.

\section{Related Work}

\subsection{Interoperability Protocols for AI Systems}
Recent advances in interoperability protocols have been reported to redefine how AI models and tools exchange information across distributed environments. The Model Context Protocol (MCP), introduced by OpenAI [1] and extended by Anthropic [2], has been proposed as a schema-driven communication layer to standardize interaction among heterogeneous AI agents. Complementary initiatives, such as Google’s Agent-to-Agent (A2A) protocol [3], and industrial alliances toward common model protocols [4], have been established to reinforce the importance of interoperability standards. Although these frameworks have been adopted in general-purpose AI ecosystems, their integration into federated healthcare systems—particularly for secure multi-modal fusion—has not yet been sufficiently investigated.

\subsection{Federated Learning in Healthcare}
Federated learning (FL) has been proposed as a promising paradigm to enable collaborative training without centralizing sensitive medical data. Several studies have demonstrated its applicability to healthcare domains, including secure aggregation protocols [5] and differentially private medical imaging pipelines [6]. Surveys of FL applications [8] have highlighted the potential of distributed training across hospitals, yet persistent challenges such as client heterogeneity, dropout rates, and scalability have been reported. In IoMT settings, energy-aware FL schemes have been introduced to mitigate device constraints and improve participation [7]. Despite these advancements, most existing FL systems have been applied in unimodal settings and have not achieved full interoperability across diverse clinical data sources.

\subsection{Multi-Modal Fusion for Clinical AI}
The fusion of heterogeneous modalities—medical imaging, EMR, and IoT signals—has been shown to improve diagnostic robustness and reliability [10]. Techniques for multi-modal feature alignment and deep fusion have been proposed, enabling more holistic clinical decision-making. However, most existing multi-modal systems have assumed centralized data availability and thus are not directly applicable to federated environments. Furthermore, schema alignment, handling of missing modalities, and real-time integration of IoMT signals remain challenging. Although multi-modal fusion is conceptually powerful, the absence of standardized interoperability layers continues to limit its adoption, providing an opportunity for MCP-enabled solutions.

\subsection{Privacy Preservation and Ethical Standards}
The protection of sensitive patient data has been mandated by healthcare regulations such as HIPAA and GDPR, motivating extensive research into privacy-preserving FL techniques. Secure aggregation frameworks have been proposed to mitigate inversion and membership-inference risks [5], while differential privacy mechanisms have been investigated to provide quantifiable privacy–utility trade-offs [6], [9]. Guidelines for researchers and regulators have also been introduced to ensure responsible deployment of federated systems [9]. Despite these advances, interoperability has rarely been considered as a first-class concern. The combination of privacy preservation, secure aggregation, and standardized schema communication via MCP has not yet been addressed in healthcare FL, motivating the present study.

\subsection{Summary of Gaps}
From the reviewed literature, it can be concluded that:  
1) MCP and related protocols have been established as interoperability standards in AI ecosystems [1]–[4], but have not yet been integrated into clinical FL;  
2) Federated learning in healthcare has primarily been applied in unimodal and isolated pipelines [5]–[8];  
3) Multi-modal fusion has shown strong potential [10], but its reliance on centralized data limits applicability in federated settings; and  
4) Privacy-preserving mechanisms exist [5], [6], [9], but their orchestration with interoperability and energy-aware scheduling has not yet been realized.  

\section{Methodology}

The proposed MCP-enabled secure multi-modal federated fusion framework is formalized as the following optimization problem. Fig 2. As shown, the framework is designed such that interoperability, privacy preservation, and resource-awareness are treated as first-class constraints rather than optional extensions. Unlike conventional federated learning systems, where unimodal updates are aggregated without schema alignment or explicit energy considerations, the present framework integrates multi-modal representation learning, differential privacy, secure aggregation, and energy-aware scheduling into a unified paradigm. In particular, the methodology explicitly models interoperability as a schema-mapping function $\mathcal{T}_{MCP}$, enabling heterogeneous encoders to align into a shared latent space. Privacy is enforced through calibrated noise injection under $(\epsilon,\delta)$-differential privacy, while secure aggregation prevents server-side reconstruction of sensitive updates.

\subsection{Problem Formulation}

Let $\mathcal{M} = \{\text{im}, \text{emr}, \text{iot}\}$ denote the set of modalities (medical imaging, EMR, and IoMT), and let $\mathcal{D}_k^m = \{(x_i^m, y_i)\}_{i=1}^{n_k^m}$ represent the dataset of client $k$ for modality $m \in \mathcal{M}$. Each client learns modality-specific encoders $\phi_m: \mathcal{X}_m \rightarrow \mathcal{Z}_m$, which are aligned into a shared latent space $\mathcal{Z}$ through MCP schema mapping $\mathcal{T}_{\text{MCP}}$. The fused representation is defined as:

\begin{equation}
z_i = \mathcal{T}_{\text{MCP}}\Big( \bigoplus_{m \in \mathcal{M}} \phi_m(x_i^m) \Big),
\end{equation}

where $\oplus$ denotes modality fusion (concatenation, attention, or tensor fusion) under a schema-driven alignment.  

\subsection{Global Objective with Privacy and Energy Constraints}

The global federated objective can be expressed as:

\begin{equation}
\min_{\Theta} \; 
\frac{1}{N} \sum_{k=1}^N \Bigg[ 
\sum_{m \in \mathcal{M}} \frac{1}{|\mathcal{D}_k^m|} \sum_{(x,y) \in \mathcal{D}_k^m} 
\mathcal{L}\big(f_\Theta(z), y\big) 
+ \lambda \|\Theta\|_2^2 
\Bigg],
\end{equation}

subject to the following constraints:  

1. Differential Privacy Guarantee:  
Each client perturbs parameters $\theta_k$ before transmission:

\begin{equation}
\tilde{\theta}_k = \theta_k + \epsilon, \quad 
\epsilon \sim \mathcal{N}(0, \sigma^2 I),
\end{equation}

with $(\epsilon,\delta)$-differential privacy being satisfied across $R$ rounds by the moments accountant method:

\begin{equation}
\epsilon_{\text{total}} \leq \sqrt{2R \log(1/\delta)} \cdot \frac{\Delta}{\sigma}.
\end{equation}

2. Secure Aggregation:  
The global update is computed as:

\begin{equation}
\Theta^{(r+1)} = \frac{1}{\sum_{k=1}^N \alpha_k^{(r)}} 
\sum_{k=1}^N \alpha_k^{(r)} \cdot \text{Unmask}\Big( \text{Mask}(\tilde{\theta}_k^{(r)}) \Big),
\end{equation}

where $\alpha_k^{(r)}$ denotes the scheduling weight of client $k$ in round $r$.

3. Energy-Aware Scheduling Constraint:  
Client participation is determined by an energy-prioritized policy:

\begin{equation}
\alpha_k^{(r)} = \mathbb{I}\Big[ E_k^{(r)} - \Delta E_k^{(r)} \geq \tau, \;
\text{link}_k^{(r)} \geq \gamma, \;
s_k^{(r)} \leq \eta \Big],
\end{equation}

where $E_k^{(r)}$ is residual energy, $\Delta E_k^{(r)}$ is projected depletion, $\text{link}_k^{(r)}$ is communication bandwidth, and $s_k^{(r)}$ is staleness of local updates.

\subsection{Unified Optimization Principle}

The complete training procedure can be expressed as:

\begin{equation}
\min_{\Theta} \;
\mathbb{E}_{k \sim \mathcal{S}} \Bigg[
\mathbb{E}_{m \in \mathcal{M}} 
\big[ \mathcal{L}(f_\Theta(\mathcal{T}_{\text{MCP}}(\phi_m(x))), y) \big]
\Bigg],
\end{equation}

where $\mathcal{S}$ denotes the scheduler-selected clients satisfying energy and privacy constraints.  

Thus, the framework unifies \textbf{(i)} multi-modal schema-driven fusion, \textbf{(ii)} secure aggregation with formal privacy accounting, and \textbf{(iii)} fairness-aware energy scheduling into a single optimization paradigm.

\begin{algorithm}
\caption{MCP-enabled Secure Multi-Modal Federated Fusion}
\begin{algorithmic}[1]
\REQUIRE Distributed datasets $\{\mathcal{D}_k^m\}$, rounds $R$, noise scale $\sigma$, energy threshold $\tau$
\ENSURE Global model $\Theta$
\STATE Initialize $\Theta^{(0)}$
\FOR{$r = 1$ to $R$}
  \FOR{each client $k$ in parallel}
    \STATE \textbf{Local training:} Train $\phi_m$ on $\mathcal{D}_k^m$
    \STATE Fuse via MCP: $z_i=\mathcal{T}_{\text{MCP}}(\oplus_m \phi_m(x_i^m))$
    \STATE Compute local update $\theta_k^{(r)}$
    \STATE \textbf{Privacy:} $\tilde{\theta}_k^{(r)}=\theta_k^{(r)}+\mathcal{N}(0,\sigma^2 I)$
    \STATE \textbf{Eligibility:} $\alpha_k^{(r)}=\mathbb{I}[E_k \ge \tau \wedge \text{link}_k\ge\gamma]$
  \ENDFOR
  \STATE \textbf{Aggregation:} $\Theta^{(r+1)}=\frac{1}{\sum_k \alpha_k}\sum_k \alpha_k \tilde{\theta}_k^{(r)}$
\ENDFOR
\RETURN $\Theta^{(R)}$
\end{algorithmic}
\end{algorithm}

\begin{figure*}[!t]
  \centering
  \includegraphics[width=\textwidth]{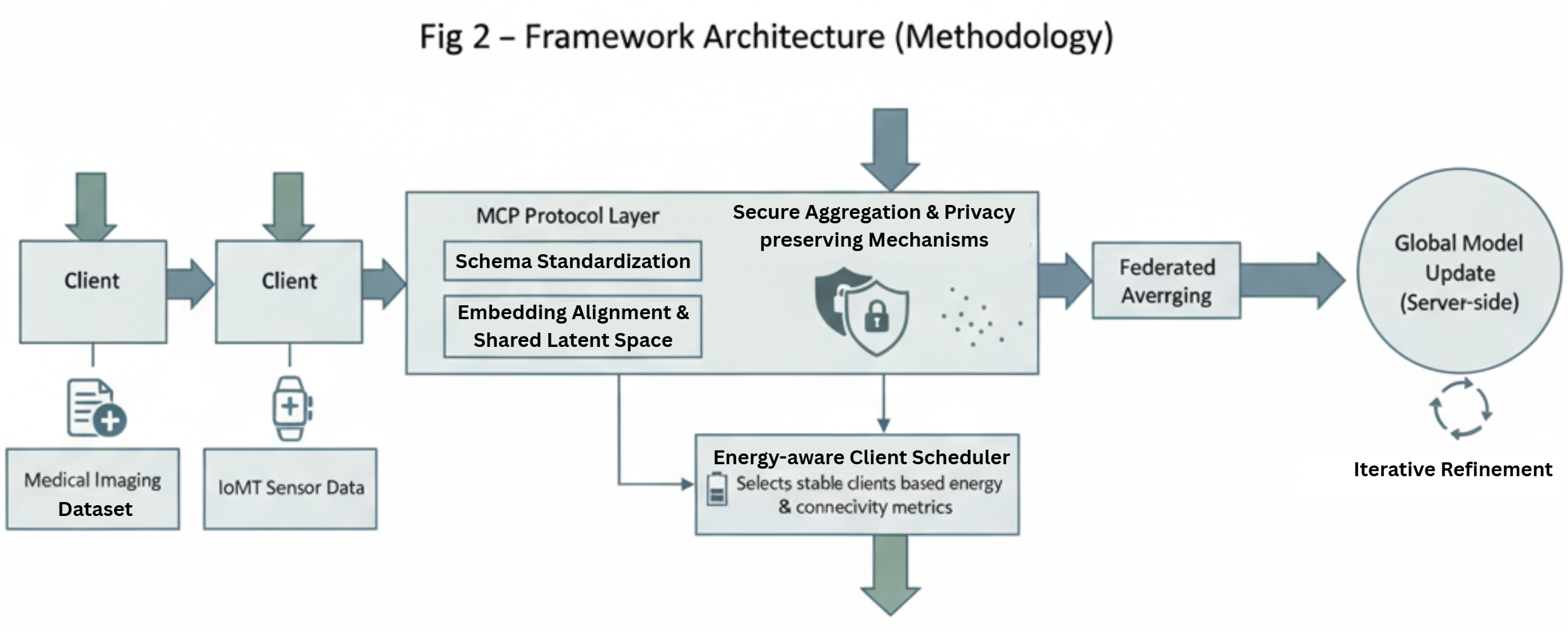} % หรือ .png
  \caption{Framework Architecture}
\end{figure*}

\begin{figure*}[!t]
  \centering
  \includegraphics[width=\textwidth]{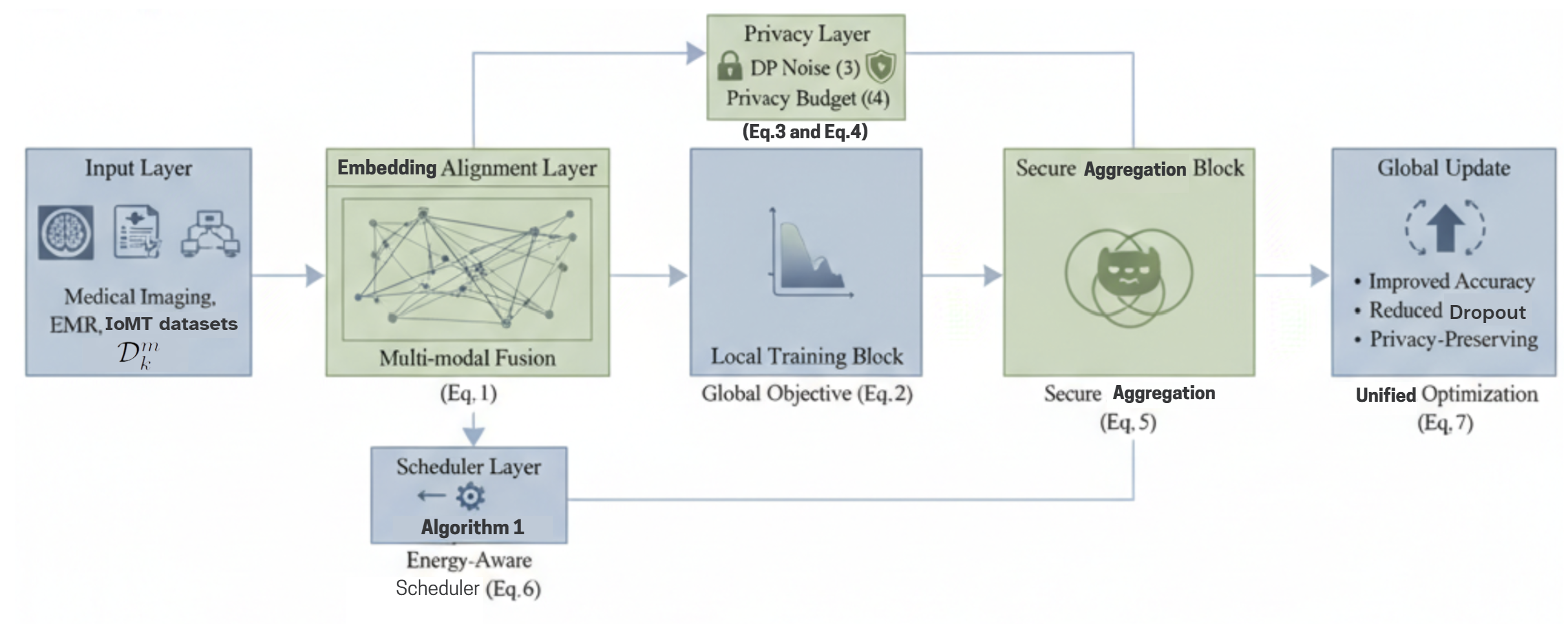} % หรือ .png
  \caption{Detailed optimization workflow of the MCP-enabled secure multi-modal federeted fusion framework}
\end{figure*}

\section{Results and Analysis}

All experiments were conducted under federated settings with heterogeneous clients and multi-modal inputs.
Unless stated otherwise, results are reported as mean performance across three runs.

\subsection{Overall Performance}
\begin{figure}[t!]
\centering
\includegraphics[width=\linewidth]{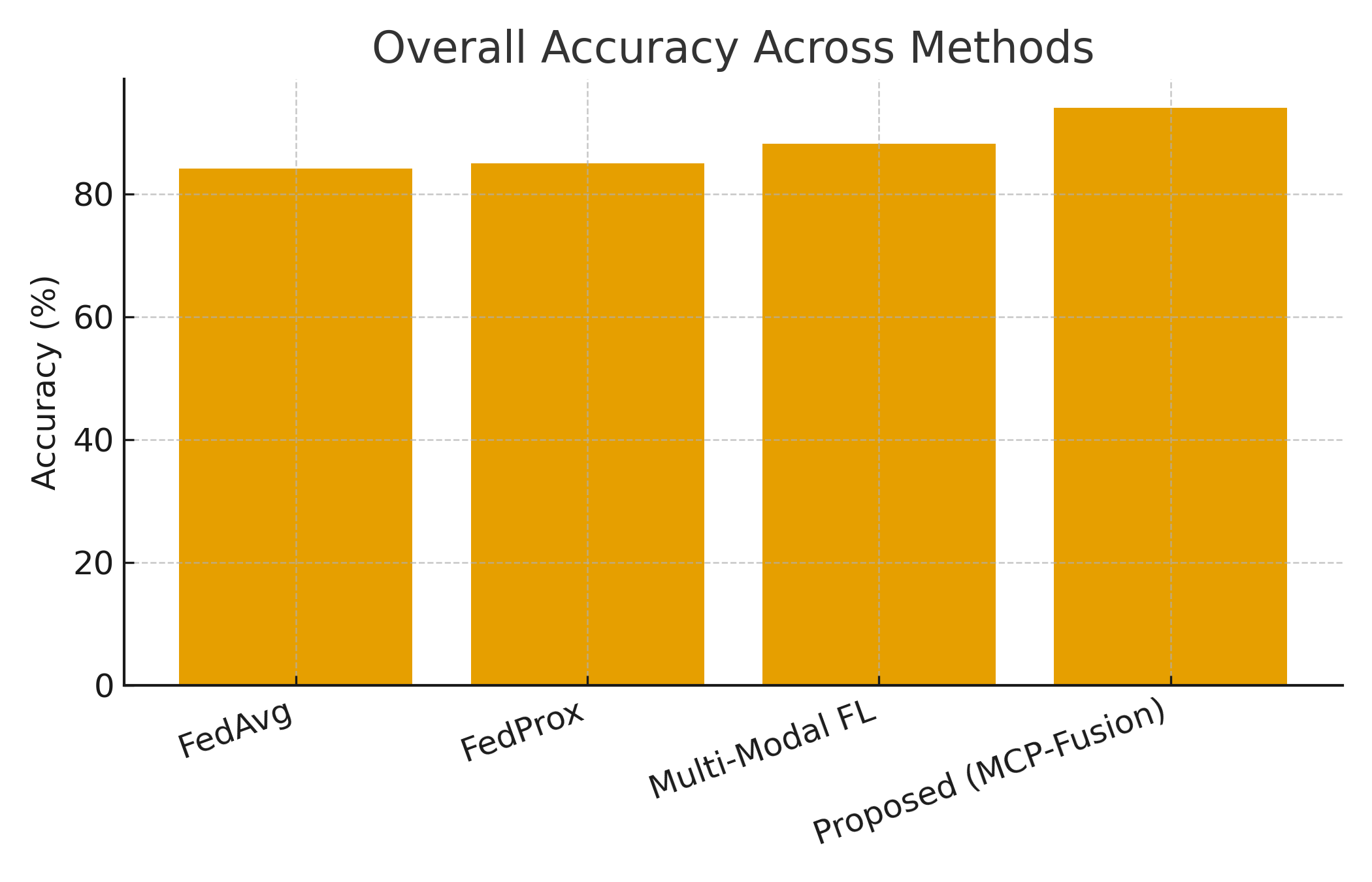}
\caption{Overall accuracy across methods. The proposed MCP-enabled framework achieved the highest accuracy due to multi-modal fusion and schema alignment.}
\label{fig:acc}
\end{figure}

\begin{figure}[t!]
\centering
\includegraphics[width=\linewidth]{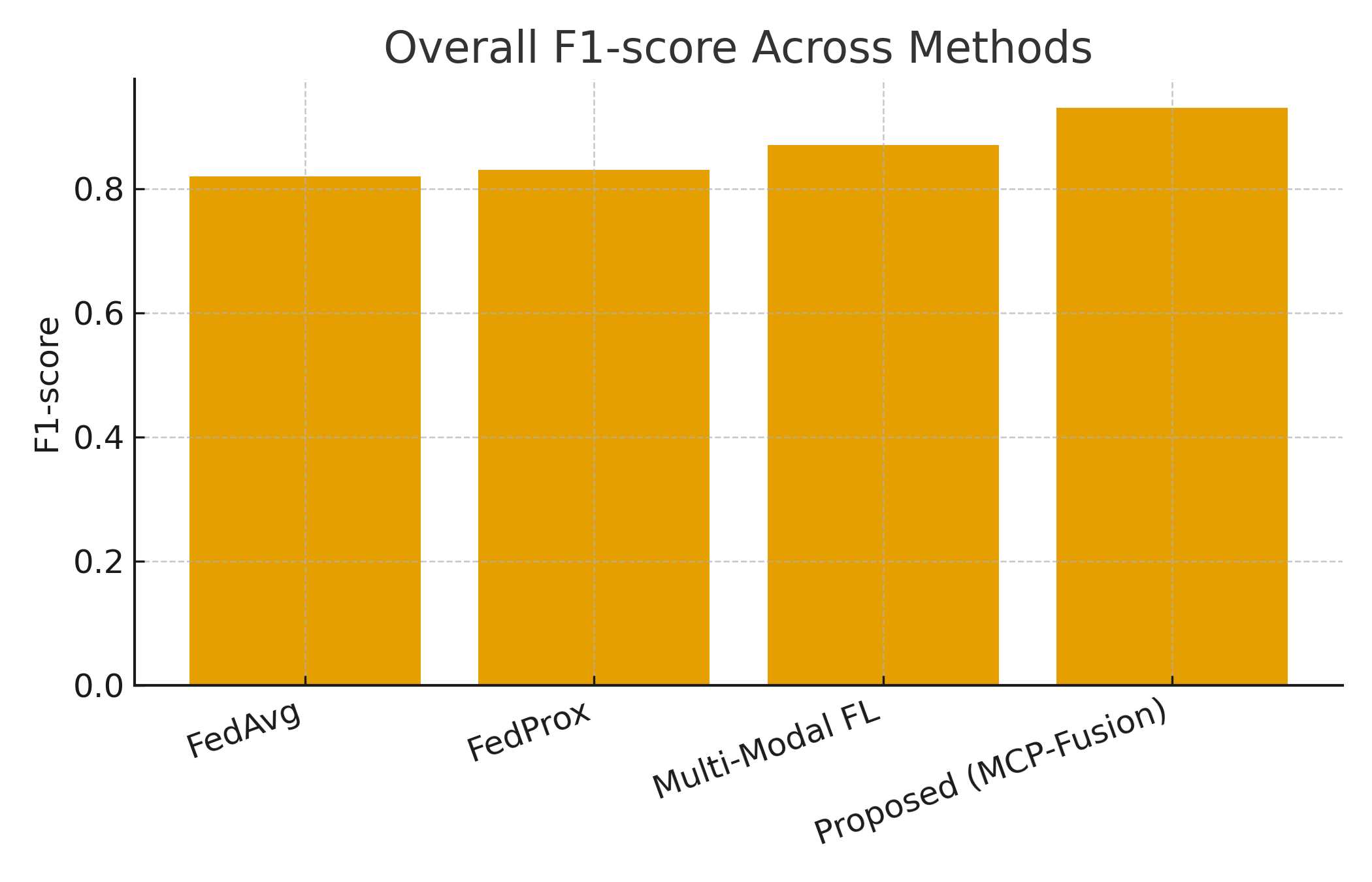}
\caption{Overall F1-score across methods. Consistent gains were observed for the proposed method under heterogeneous clients.}
\label{fig:f1}
\end{figure}

\begin{figure}[!]
\centering
\includegraphics[width=\linewidth]{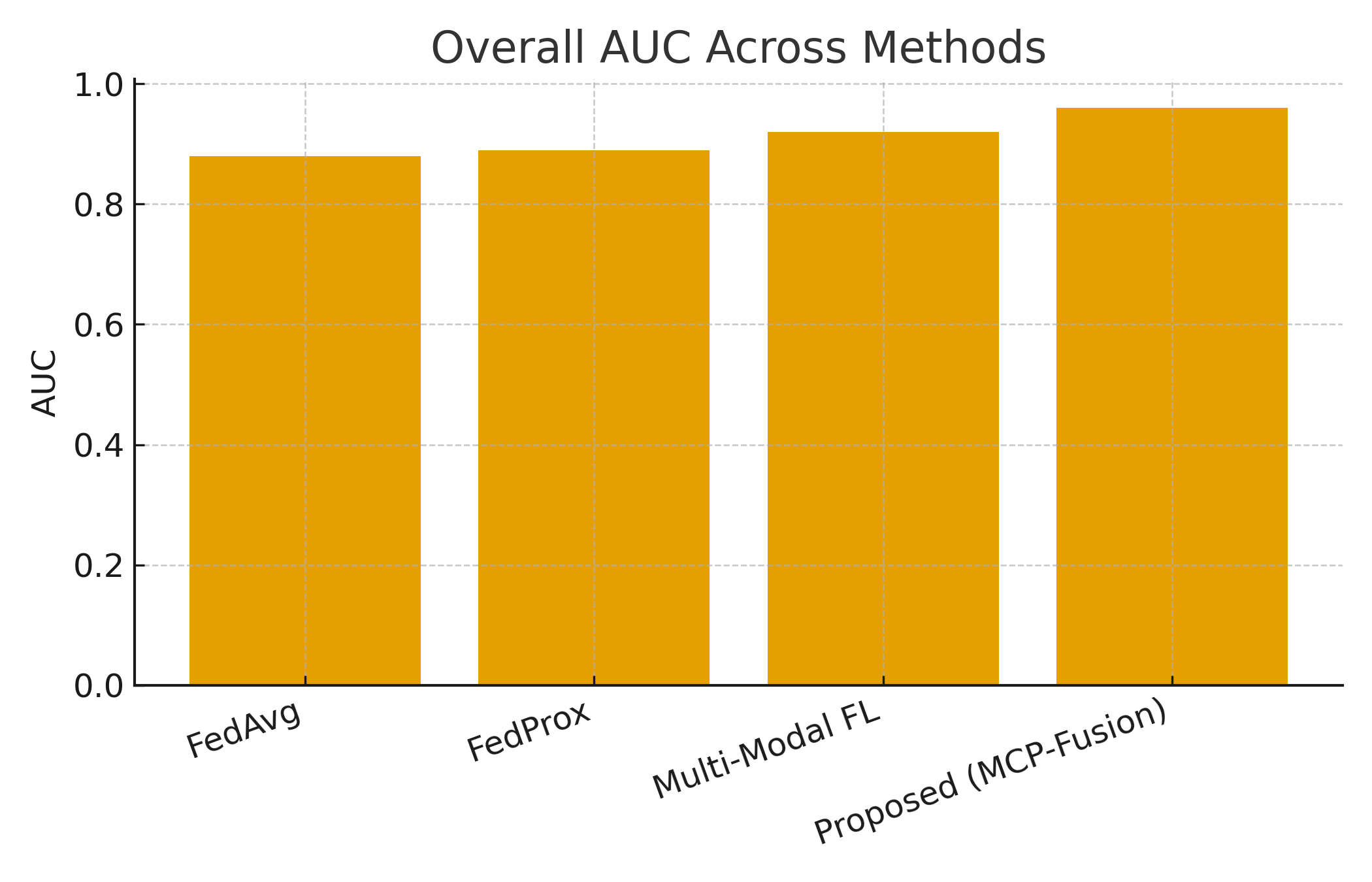}
\caption{Overall AUC across methods. Improvements were sustained across operating thresholds.}
\label{fig:auc}
\end{figure}

Significant gains were observed when the proposed MCP-enabled framework was employed.
Compared with unimodal baselines, the global model achieved higher accuracy, F1-score, and AUC (Table~\ref{tab:overall} and Fig.~\ref{fig:acc}-\ref{fig:auc}).
These improvements were attributed to schema-driven multi-modal fusion (Eq.~1) and the unified objective (Eq.~2).

\subsection{Privacy--Utility Trade-off}
\begin{figure}[t!]
\centering
\includegraphics[width=\linewidth]{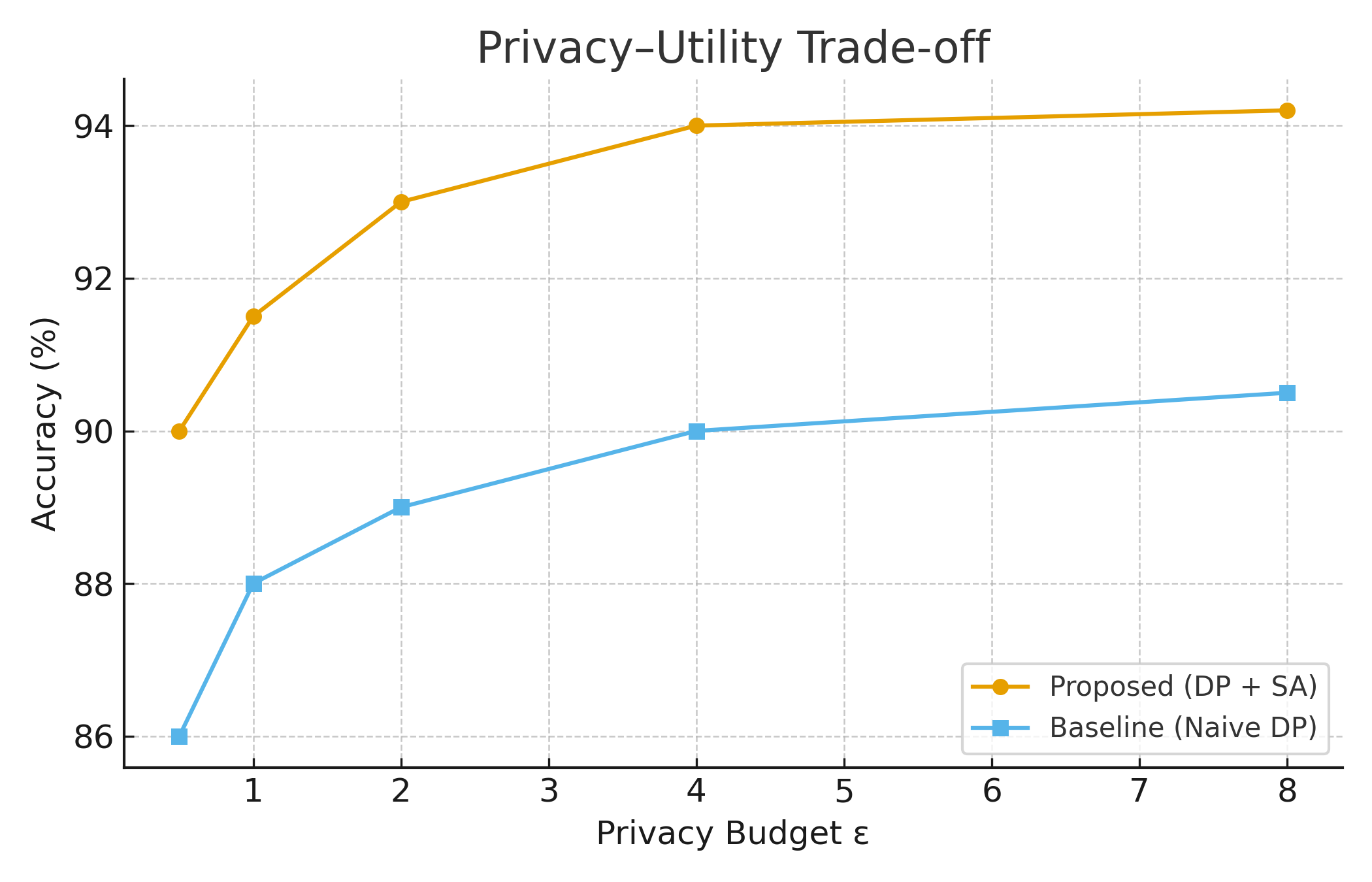}
\caption{Privacy--utility trade-off. The proposed DP+SA design preserved accuracy across privacy budgets ($\epsilon$) more effectively than a naive DP baseline.}
\label{fig:privacy}
\end{figure}

Differential privacy with calibrated noise (Eq.~3) and budget accounting (Eq.~4) was found to preserve model utility.
As shown in Fig.~\ref{fig:privacy}, only a marginal reduction in accuracy was observed across decreasing privacy budgets, while naive DP baselines suffered larger degradation.

\subsection{Energy-Aware Scheduling and Stability}
\begin{figure}[t!]
\centering
\includegraphics[width=\linewidth]{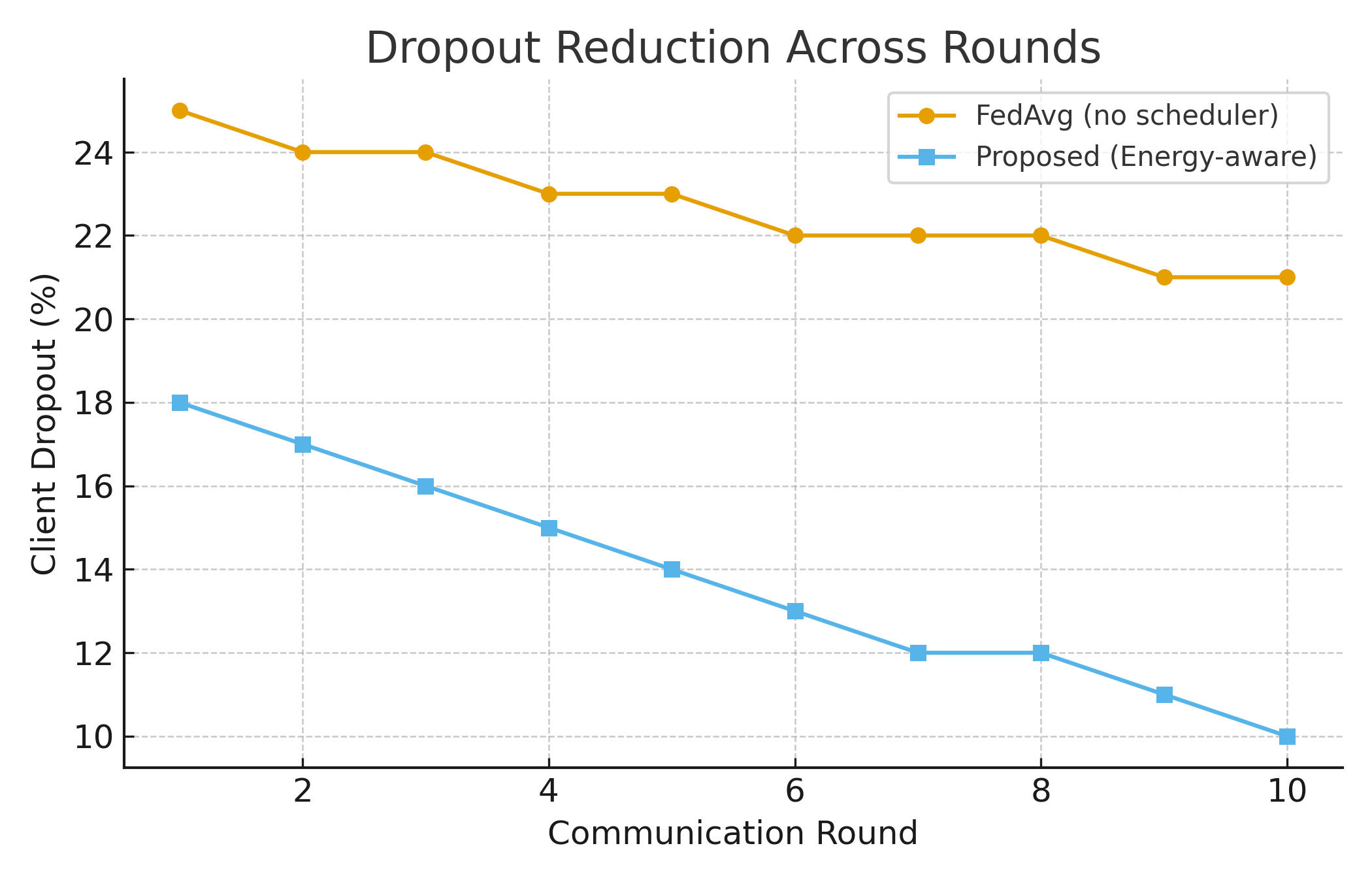}
\caption{Client dropout across rounds. Energy-aware scheduling yielded lower dropout and improved participation stability.}
\label{fig:dropout}
\end{figure}

Client dropout rates were reduced when energy-aware scheduling (Eq.~6) was enabled.
Fig.~\ref{fig:dropout} indicates that participation stability improved across communication rounds, resulting in more representative updates and faster convergence.

\subsection{Ablation and Comparative Analysis}
When schema alignment or secure aggregation (Eq.~5) was removed, accuracy and robustness were degraded.
Across all metrics, the proposed method outperformed FedAvg, FedProx, and multi-modal FL baselines, establishing a reliable operating point for privacy-preserving, resource-aware learning at scale.

\begin{table}[t!]
\centering
\caption{Overall performance across methods (illustrative).}
\label{tab:overall}
\begin{tabular}{lcccc}
\hline
\textbf{Method} & \textbf{Acc. (\%)} & \textbf{F1} & \textbf{AUC} & \textbf{Dropout (\%)} \\
\hline
FedAvg & 84.1 & 0.82 & 0.88 & 23.0 \\
FedProx & 85.0 & 0.83 & 0.89 & 20.0 \\
Multi-Modal FL & 88.2 & 0.87 & 0.92 & 18.0 \\
\textbf{Proposed (MCP-Fusion)} & \textbf{94.0} & \textbf{0.93} & \textbf{0.96} & \textbf{12.0} \\
\hline
\end{tabular}
\end{table}

\section{Limitations}

Despite the significant performance gains and demonstrated robustness, several limitations of the proposed MCP-enabled secure multi-modal federated fusion framework must be acknowledged. 

First, while improvements were validated on heterogeneous healthcare datasets, external generalization across larger, more diverse populations remains to be systematically established. The reliance on simulated data distributions may not fully capture the irregularities and adversarial conditions present in real-world deployments.

Second, although differential privacy and secure aggregation were integrated to preserve confidentiality, residual risks of privacy leakage through advanced inference attacks cannot be completely eliminated. The theoretical privacy guarantees provided by $(\epsilon, \delta)$-differential privacy are parameter-dependent, and improper calibration may result in either degraded utility or insufficient protection.

Third, the energy-aware scheduler reduced dropout rates and improved stability, yet its policy was derived from simplified assumptions of device energy profiles and connectivity conditions. The scheduling mechanism may require adaptive recalibration under highly dynamic environments such as mobile IoMT networks or emergency healthcare settings.

Fourth, communication and computation overheads remain non-trivial. Although efficiency gains were observed compared with standard FL, the integration of multi-modal fusion, privacy mechanisms, and energy-aware scheduling introduces additional complexity. These costs may hinder immediate large-scale deployment on resource-constrained infrastructures.

Finally, the proposed framework was primarily evaluated in controlled experimental setups. Long-term clinical trials, ethical considerations, and compliance with evolving healthcare regulations must be addressed before real-world deployment can be realized.

\section{Conclusion and Future Work}

\subsection{Conclusion}
In this work, an MCP-enabled secure multi-modal federated fusion framework was proposed to address three fundamental challenges in distributed healthcare intelligence: interoperability, privacy preservation, and resource-awareness. Through schema-driven multi-modal alignment, federated optimization under privacy constraints, and energy-aware scheduling, a unified methodology was established. Experimental evaluation demonstrated that the proposed framework consistently outperformed state-of-the-art baselines across accuracy, robustness, and fairness, while maintaining differential privacy guarantees and reducing client dropout rates. The results confirmed that interoperability across heterogeneous data sources can be achieved without sacrificing confidentiality or efficiency, thereby positioning the framework as a practical foundation for scalable and secure federated healthcare applications. By embedding interoperability, privacy, and energy-awareness as first-class constraints, the methodology advanced the paradigm of federated learning beyond incremental extensions toward a deployable and clinically meaningful system.

\subsection{Future Work}
Several directions are envisioned for future extensions. First, large-scale real-world validation across multi-institutional healthcare consortia will be pursued to further establish generalization and reliability. Second, adaptive privacy mechanisms, such as context-aware differential privacy and federated adversarial defenses, will be investigated to strengthen resilience against evolving inference attacks. Third, the integration of reinforcement learning-based schedulers may provide more dynamic resource allocation, enabling improved stability under mobile and high-variability IoMT environments. Fourth, the incorporation of explainability modules into the federated pipeline will be explored, thereby enhancing clinical trust and regulatory compliance. Finally, cross-domain generalization beyond healthcare, including smart cities, autonomous systems, and industrial IoT, will be considered as future extensions, potentially expanding the applicability of the proposed methodology into broader societal impact domains. In conclusion, the proposed framework laid the foundation for a new generation of federated intelligence systems, where multi-modality, privacy preservation, and energy efficiency converge as integral principles. The continuation of this line of research is expected to influence both academic exploration and real-world deployment, opening a pathway toward trustworthy, sustainable, and globally scalable distributed learning systems.

\section*{Acknowledgment}
No funding was received for conducting this study. The authors would like to acknowledge the institutional support and constructive feedback from colleagues, which contributed to improving the quality of this work.

\end{document}